# Focusing through turbid media by polarization modulation

Jongchan Park[1], Jung-Hoon Park[1,2], Hyeonseung Yu[1], and YongKeun Park[1*]

[1]*Department of Physics, Korea Advanced Institute of Science and Technology, Daejeon 305-701, South Korea*
[2]*Current address: Howard Hughes Medical Institute, Janelia Research Campus 19700 Helix Drive, Ashburn, VA 20147, USA*
*\*Corresponding author: yk.park@kaist.ac.kr*

We demonstrate that polarization modulation of an illumination beam can effectively control the spatial profile of the light transmitted through turbid media. Since the transmitted electric fields are completely mingled in turbid media, polarization states of an illumination beam can be effectively used to control the propagation of light through turbid media. Numerical simulations were performed which agree with experimental results obtained using a commercial in-plane switching liquid-crystal display for modulating the input polarization states.

When coherent waves propagate through highly disordered media, the output wavefront generates random speckle patterns due to multiple scattering events. As this phenomena looks like a stochastic random process, multiple scattering was regarded as a major obstacle in the field of optics until recently. However, multiple scattering events are deterministic for each specific input wavefront and the distribution of disorder. Thus, the randomness of highly scattering media, which has millions of degrees of freedom, can be utilized for functional advantages rather than trying to be avoided [1-3]. The generation of a focus behind a disordered medium is the most fundamental step for controlling random scaterring. This was demonstrated in the pioneering work by Vellekoop *et al.* [4] using the phase modulation method. In the original scheme, a phase-only spatial light modulator (SLM) was used to shape the incident wavefront for focusing behind turbid media. Another recent method is to use binary amplitude modulation where only the amplitude of the incident wavefront is controlled by a digital micromirror device [5]. The two schemes have now become the methods of choice in the field of wavefront shaping through turbid media.

Wavefront shaping is a promising method for exploiting the high degree of freedom that is brought on by randomness. Since multiple scattering events in disordered media scramble the complex electric fields, not only a single focused beam but also arbitrary wavefronts can be generated through turbid media by wavefront shaping of the incident field [2, 6-14]. Generally, phase-only SLMs are used for high efficiency shaping of the wavefront while digital micromirror devices are used for high speed applications albeit the much lower efficiency. However, the high price of such units limits the potential applications in industry.

Highly scattering media completely mix optical parameters of the propagating waves. Recently, it was reported that disordered media can be used as dynamic active wave pates [8, 10]; the polarization states of the wave transmitted from disordered media can be controlled by modulating the spatial phase profile of the incident wavefronts. If we consider time-reversal symmetry, we can also expect that in the reverse manner, the phase and amplitude of waves transmitted from turbid media can also be readily controlled by modulating the polarization states of incident waves due to the randomness of turbid media.

Here, we introduce a new method to focus light through turbid media based on polarization modulation. The polarization states, rather than amplitude or phase, of the incident wave is spatially modulated by a liquid crystal display monitor which is a much more economic and widely available modulator compared to phase-only SLMs due to mass production for various commercial displays. We first perform computer-based numerical simulations to analyze the effectiveness of the polarization modulation method and then experimentally demonstrate the method by using a commercial in-plane switching liquid crystal display (IPS-LCD) [15].

The transmission of light through optical elements can be effectively described by using the transmission matrix (TM), which describes the complex linear relationship between incident and transmitted light [16, 17]. When the area of the turbid medium is spatially divided into the size of a diffraction limit spot, each spot can be considered as an individual channel for orthogonal optical modes. As the propagation of light through optical elements modulate the phase as well as the amplitude, the TM consist of complex entries describing the complex optical relationships between each input and output channels. The TM allows us to predict the relation between the input and output fields even if we cannot follow the exact optical trajectories of waves propagating in complex media.

Recently, the vector TM was measured which took into account the higher degree of freedom that can be added by relating both complex electric fields and polarization states of incident and outgoing waves [18, 19]. The input and output channels of the vector TM describe the optical modes of two orthogonal polarization states, thus when a wavefront is incident on the medium the optical responses in terms of polarization states of the single channel in real space can be described by the linear combination of the output signals from two corresponding orthogonal channels. The vector TM can be described as shown below [18]:

$$\begin{bmatrix} E_{out,1}^{x} \\ \vdots \\ E_{out,N}^{x} \\ E_{out,1}^{y} \\ \vdots \\ E_{out,N}^{y} \end{bmatrix} = \begin{bmatrix} t_{11}^{xx} & \cdots & t_{1M}^{xx} & t_{11}^{xy} & \cdots & t_{1M}^{xy} \\ \vdots & \ddots & \vdots & \vdots & \ddots & \vdots \\ t_{1N}^{xx} & \cdots & t_{NM}^{xx} & t_{1N}^{xy} & \cdots & t_{NM}^{xy} \\ t_{11}^{yx} & \cdots & t_{1M}^{yx} & t_{11}^{yy} & \cdots & t_{1M}^{yy} \\ \vdots & \ddots & \vdots & \vdots & \ddots & \vdots \\ t_{1N}^{yx} & \cdots & t_{NM}^{yx} & t_{1N}^{yy} & \cdots & t_{NM}^{yy} \end{bmatrix} \begin{bmatrix} E_{in,1}^{x} \\ \vdots \\ E_{in,M}^{x} \\ E_{in,1}^{y} \\ \vdots \\ E_{in,M}^{y} \end{bmatrix}, \quad (1)$$

where the complex valued entry $t_{lm}^{ij}$ relates the $j$th axis component of the $m$th input channel and $i$th axis component of $n$th output channel (Here, the 1st and 2nd axis corresponds to the $x$- and $y$-axis components, respectively).

Using the vector TM, we investigated the output field responses when the polarization states of the incident fields are modulated. We consider the situation where an iterative feed-back algorithm is used to enhance the intensity signal from a single output channel, thus generating a focused beam behind the turbid medium. The intensity enhancement factor of the generated focus, $\eta$ is defined as, $\eta = I_f / I_{bg}$, where $I_f$ is the intensity of the target output channel and $I_{bg}$ is the averaged intensity of all other output channels excluding the target channel from the final output wavefront.

To evaluate the enhancement factor of our method, we performed a numerical simulation of optical wave propagation in a highly disordered medium. The TM that we used was generated by using random complex entries that have a Gaussian distribution due to the Central Limit Theorem; although the entries of scattering matrices of real turbid media have weak correlations, we can safely assume that it is valid to use uncorrelated entries as we cannot observe open channels in our current experimental regime [17]. Using the vector TM, input and output channels of the TM were divided into two sets of orthogonal polarization bases (Here, we divided them into $x$- and $y$-polarization components). Thus, vector summation of the two corresponding orthogonal polarization signals describes polarization states of a single desired channel in real space.

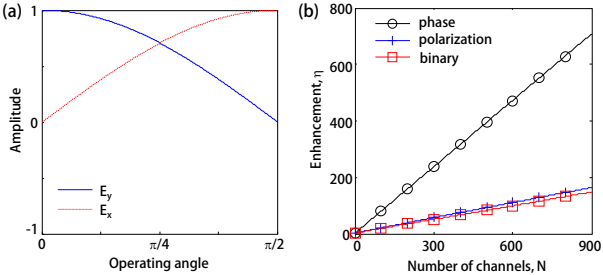

Fig. 1. (a) Amplitude modulation profiles of incident orthogonal channels. (b) Intensity enhancement versus number of incident channels for each methods.

In our numerical simulation, perfect polarization state modulation using an ideal IPS-LCD was assumed. In this condition, anisotropic liquid crystal molecules of IPS-LCD pixels rotate in the in-plane direction, where each individual pixel plays the same role as a conventional half-wave plate. For a linearly polarized incident wave, the phases of the two orthogonal polarization components of a wave remains unchanged while the amplitudes of each component are modulated due to the rotation of polarization direction. The amplitude modulation profile of each corresponding orthogonal channel for the simulation was set as shown in Fig. 1(a). The graph illustrates the amplitude modulation of each orthogonal component of the incident field as a function of the rotation angle of the liquid crystal molecules. Initially, the oscillating orientation of the linearly polarized incident wave was set parallel to the $y$-axis which rotates proportionally with the liquid crystal rotation up to a maximum angle of 90º which corresponds to the polarization rotation of the input wavefront to the orthogonal $x$-axis (corresponding to a dark and bright pixel in a conventional LCD monitor).

The acquired intensity enhancements of the polarization modulation method is compared with those of using other previous methods [Fig. 1(b) and Table 1]; numerical simulations of phase modulation, binary amplitude modulation and polarization modulation were performed under the same condition. A total of 1000 numerical simulations were done and ensemble averaged for each method. For each simulation, the target output channel for intensity enhancement was randomly chosen while the transmission matrices were also randomly generated for every single simulation. The transmission matrices were generated for 900 input and output channels. The acquired enhancement factor of the phase modulation and the binary amplitude modulation from our computer-based simulations are consistent with previously studied analytical solutions [4, 5], showing the validity of our simulations. The enhancement factor per controlled number of input channels of the polarization modulation method is 0.182 which is slightly better than the binary amplitude modulation method due to more efficient use of the input energy.

Table 1. Intensity enhancement of each method

| ethod | Enhancement ($\eta/N$) | # of channels |
|---|---|---|
| Phase | $0.707 \pm 0.001$ | 900 |
| Binary | $0.164 \pm 0.001$ | 900 |
| Polarization | $0.182 \pm 0.001$ | 900 |
| IPS-LCD (sim.) | $0.108 \pm 0.001$ | 900 |
| IPS-LCD (exp.) | $0.032 \pm 0.009$ | 900 |

Based on the positive results from the numerical simulations, we also demonstrate the experimental realization of the polarization modulation method and successfully generate a focus behind a highly turbid medium, which is made of white paint (Pingo General, Noroo Paint, South Korea), using a commercial IPS-LCD (FLATRON 20EA34TQ, LG electronics, South Korea).

The experimental scheme is similar to the method developed by Vellekoop *et al.* [4]. In the optical setup, light from a He-Ne laser ($\lambda$ = 633 nm, 5.0 mW, Thorlabs, USA) is spatially filtered and expanded by a lens pair. A polarizer and a waveplate are placed in order to control the polarization state of the incident wave. An iris is placed in front of the IPS-LCD to illuminate a total of 3600 pixels of the IPS-LCD. Polarization sheets attached to the LCD panel were removed before experiments.

The thickness and transport mean free path (TMFP) of the white paint sample was measured as 19.6 ± 1.1 μm and 1.2 ± 0.2 μm, respectively. TMFP was obtained by measuring the total transmittance of light as a function of sample thickness [20, 21].

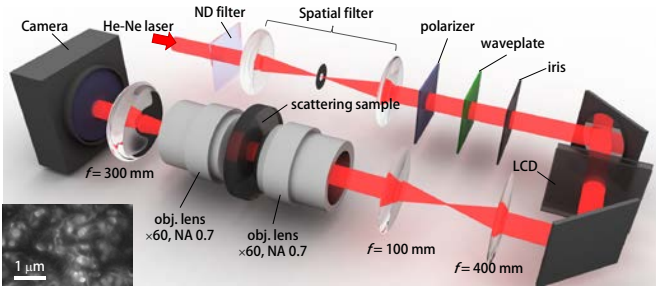

Fig. 2. Experimental set-up for generating focusing behind a scattering sample using commercial IPS-LCD. Polarization states of incident waves are modulated by the IPS-LCD. (*inset*) Surface image of white paint sample obtained by scanning electron microscopy.

The polarization state of an incident wave is spatially modulated by the IPS-LCD, and then focused into the white paint turbid layer. The IPS-LCD is conjugated to the Fourier plane of the scattering sample; thus, each spatial mode of the wave transmitted from the IPS-LCD is converted into an angular mode in order to give homogeneous energy distribution behind the specific target position of the scattering sample. The speckle field behind the scattering sample is collected by an objective lens (LUCPLFL, ×60, NA = 0.7, Olympus, USA) and relayed through a 4-*f* telescopic imaging system, with the total lateral magnification of ×100. The speckle field image is observed by a CMOS camera (C11440-22C, Hamamatsu, Japan) which has 2048×2048 pixels with a 6.5-μm pixel size. In our measurements, an array of 2×2 pixels was binned into a super pixel to reduce the effect of observation errors.

To implement our method, a continuous sequential algorithm with positive feedback is adopted for optimizing the polarization modulation. Four pixels in the IPS-LCD form a superpixel to control a single spatial mode (an input channel) in order to enhance the signal to noise ratio (SNR). Intensity of the target position behind the sample is recorded as a function of applied bias on each superpixel of the IPS-LCD. Then, the bias value which gives the maximum intensity at the target position is found and recorded. The final linear summation of the output waves resulting from the 900 optimized input channels result in the optimized focus at the output targeted position.

Figure 3 shows the intensity enhancement profile of the generated focus (blue solid line) as a function of optimized channel number. A total of 900 channels were used to modulate the polarization states of the incident wave. The resulting enhancement factor from our experiment was 28.8, which is lower than the ideal enhancement factor.

The major reason of the lower enhancement factor from the experiment is from the imperfect polarization modulation of the commercial IPS-LCD. Unlike the ideal in-plane polarization modulation mode which allows perfect transition between orthogonal states, the polarization is found to rotate only up to 45° using the commercial LCD which is significantly lower.

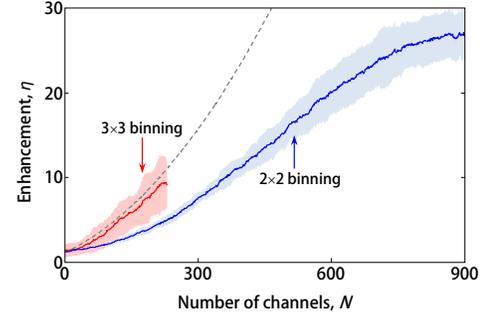

Fig 3. Intensity enhancement versus the number of controlled channels in the IPS-LCD. Blue and red solid lines, ensemble averages of measured enhancements using 230 and 900 channels, respectively. Shaded area, standard deviation of measurements. Dotted line, computer-based numerical simulation for imperfect polarization modulation of IPS-LCD. Number of measurements, $n = 10$.

To take this instrumental artifact into account, both the amplitude and phase modulation profile of a single pixel in the IPS-LCD was measured exactly as a function of applied bias [22]. Polarization sensitive digital holographic microscopy was used to measure the full Jones matrix components which describe the complex optical responses for arbitrary incident waves. Figures 4(a) and (b) show the amplitude and phase modulation profile of the measured Jones matrix components of a pixel in the IPS-LCD. $J_{ij}$ ($i,j = 1,2$) denotes the complex optical responses of the outgoing electric field of the $i^{th}$ axis component to the incident electric field of the $j^{th}$ axis component.

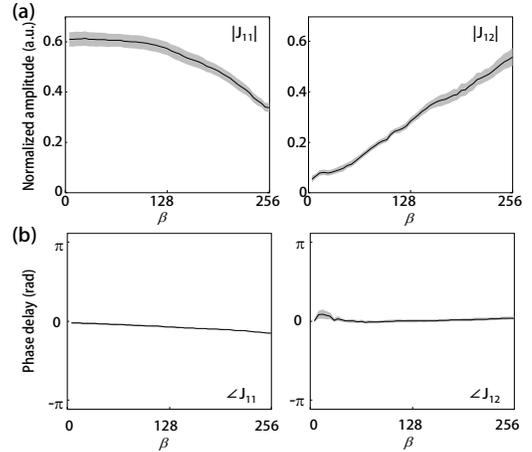

Fig. 4. Amplitude and phase modulation profiles of Jones matrix components of a pixel of the IPS-LCD as a function of applied bias. Solid lines, averaged value, Shaded areas, standard error, $n = 5$. Modified from Ref. [22].

Numerical simulations were again performed using the measured complex optical response of the IPS-LCD to calculate the enhancement factor taking into account the imperfect polarization modulation [the dotted line in Fig. 3]. However, the experimentally acquired enhancement factor is still lower than that of simulations by a factor of approximately 3. Another factor that causes the lower enhancement is that in the numerical simulations, the turbid medium and the optical setup is considered to be perfectly stationary whereas in real experiments, there are always drift and noise which change the propagation of waves in time [4, 23]. In the experimental demonstration, each measurement takes approximately 2,500 s, while the decorrelation time of our setup was measured to be 5,000 s. The SNR inherent in real life experiments also results in saturation of the enhancement differing from numerical simulations [24]. The standard deviation in the intensity fluctuation of the speckle field measured over time at a single detection point was 0.6%, which is significantly high compared to the signal level introduced by modulating a single channel of LCD.

We attribute this considerable discrepancy between the experimental and numerical simulation results to the limited SNR. When we enhanced the SNR by binning 3×3 adjacent pixels into larger superpixels, the intensity enhancement of our experimental result becomes well consistent with the prediction by numerical simulations, as shown in Fig. 3 (the red lines). The relative modulation of the input beam is redistributed among a smaller number of controlled channels resulting in an increased signal level and decreased measurement time.

To verify the universality of our method, we generate multiple foci behind the turbid media. Figure 5(c) shows the best result of the intensity profile of three generated foci. The average intensity enhancement factors of each foci was approximately 12, which is consistent with our expectation where the energy would be equally distributed on average with respect to the case where a single focus is generated using the same number of input channels.

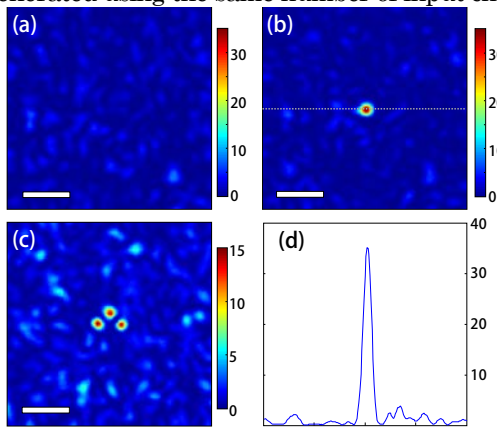

Fig. 5. Intensity profiles of transmitted light through white paint scattering sample. (a) Transmitted light with unmodulated incident wave. (b) Transmitted light after optimization for generating focus and (c) multiple foci. (d) Intensity profile of the generated focus in (b) along the white dotted line. Scale bar, 3 μm.

In summary, we proposed and experimentally demonstrated a new method to control the spatial distribution of light transmitting through turbid medium, modulating the polarization states of individual input channels instead of the amplitude or phase. The polarization modulation method enables us to utilize commercial ISP-LCD as a SLM which is economic compared to the use of phase-only SLMs. Due to the millions of degrees of freedom inherent in highly scattering media, we expect that arbitrary output wavefronts can also be generated by using our method instead of generating only a focus.

This work was supported by KAIST, the Korean Ministry of Education, Science and Technology (MEST), and the National Research Foundation (2012R1A1A1009082, 2013K1A3A1A09076135, 2013M3C1A3063046, 2012-M3C1A1-048860).